\documentclass[conference]{ACM_SIG}
\usepackage{amsmath,amssymb,latexsym}
\usepackage{graphicx}
\usepackage{subfigure}
\usepackage{cite}
\usepackage{epsfig}

\begin{document}

\title{A Coverage Theory of Bistatic Radar Networks: \\ Worst-Case Intrusion Path and Optimal Deployment}

\maketitle


\begin{abstract}

Continuing advances in radar technology are driven both by new application regimes and technological innovations. A family of
applications of increasing importance is detection of a mobile target intruding into a protected area, and one type of radar system
architecture potentially well suited to this type of application entails a network of cooperating radars. In this paper,
we study optimal radar deployment for intrusion detection, with focus on network coverage. In contrast to the disk-based
sensing model in a traditional sensor network, \emph{the detection range of a bistatic radar depends on the locations of both the radar transmitter and radar receiver, and is characterized by Cassini ovals}. Furthermore, in a network with multiple radar transmitters and receivers, since any pair of transmitter and receiver can potentially form a bistatic radar, \emph{the detection ranges of different bistatic radars are coupled and the corresponding network coverage is intimately related to the locations of all transmitters and receivers,} making the optimal deployment design highly
non-trivial. Clearly, the detectability of an intruder depends on the highest SNR received by all possible bistatic radars. We
focus on the worst-case intrusion detectability, i.e., the minimum possible detectability along all possible intrusion paths. Although it is plausible to deploy radars on a shortest line segment across the field, it is not always optimal in general, which we illustrate via counter-examples. We then present a sufficient condition on the field geometry for the optimality of shortest line deployment to hold. Further, we quantify the
local structure of detectability corresponding to a given deployment order and spacings of radar transmitters and receivers, building on which we characterize the optimal deployment to maximize the worst-case intrusion detectability. Our results show that the optimal deployment locations exhibit a balanced structure. We also develop a polynomial-time approximation algorithm for characterizing the worse-case intrusion path for any given locations of radars under random deployment.





\end{abstract}

\begin{keywords}
bistatic radar networks, network coverage, optimal deployment, worst-case intrusion path
\end{keywords}

\newtheorem{thm}{\bf Theorem}
\newtheorem{lm}{\bf Lemma}
\newtheorem{df}{\bf Definition}
\newtheorem{as}{\bf Assumption}
\newtheorem{cor}{\bf Corollary}
\newtheorem{pb}{\bf Problem}
\newtheorem{pp}{\bf Property}

\section{Introduction}

\subsection{Motivation}

\begin{figure}[t]
\centering
\includegraphics[width=0.40\textwidth]{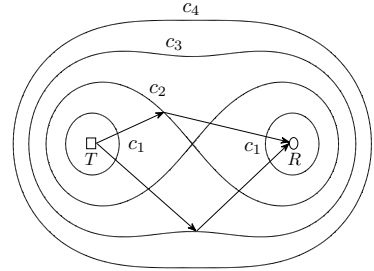}
\vspace{-1cm}
\caption{Bistatic radar SNR contours as Cassini ovals with foci at radar transmitter $T$ and radar receiver $R$
for distance products: $c_1 < c_2 < c_3 < c_4$.} \label{fg:cassini}
\end{figure}

An active radar system consists of a collection of transmitters and receivers in which transmitters emit RF signals
and the receivers capture signals resulting from scattering of the transmitted signals reflected by objects of interest (targets).
A \textit{monostatic radar} is one in which a single transmitter and single receiver are collocated.
A \textit{bistatic radar} is also comprised of a single transmit-receive pair, but the transmitter and receiver are at different locations. \textit{Multistatic radar} refers to a configuration that consists of multiple receivers at different locations.
Although physical-layer issues in radar have been extensively studied \cite{Willis}, very limited attention has been
paid to radar network design \cite{Baker03,Paolini08,Bartoletti10}. Notably, \cite{Baker03} considered a network of monostatic
radars, and \cite{Paolini08,Bartoletti10} studied a multistatic radar with one transmitter. We are thus motivated to consider a
network of multiple radar transmitters and receivers where any pair of transmitter and receiver can form a bistatic radar.
Unlike traditional passive sensors, \emph{one salient feature of radar sensing is that it actively emits signals to illuminate the
target.} Further, radar technology provides superior penetration capability by using radio waves. Worth mentioning is that the
cost of radars can be much higher than traditional sensors.


The problem of detecting intrusion into a protected area, across a border or security perimeter, is of increasing
interest for numerous applications, and has great potential in several regimes of growing importance, such as border monitoring,
industrial and airport security, and drug interdiction. Worst-case coverage \cite{Meguerdichian01INFOCOM} is a common criterion of efficacy in intrusion detection that has been applied extensively in the context of sensor networks \cite{Meguerdichian01MOBICOM,Li03TC,Gau06,Lee10}.
It is intimately related to the minimum possible detectability of an intruder (target) traversing a monitored field. In a bistatic radar network, the worst-case coverage is quite different and more difficult to quantify than traditional sensor networks, because 1) departing from the disk-based sensing range of a traditional sensor, the detection range of a bistatic radar depends on the locations of both radar transmitter and receiver and is characterized by the Cassini oval (in scenarios where intruder detectability is dominated by SNR). Formally, a Cassini oval is a locus of points for which the distances to two fixed points (foci) have a constant product (as illustrated in Figure~\ref{fg:cassini}); 2) the sensing ranges of different bistatic radars are coupled with each other, since each transmitter (or receiver) can pair with other receivers (or transmitters) to form multiple bistatic radars, indicating that its location would impact multiple bistatic radars.


\subsection{Summary of Main Contributions}

\begin{itemize}

\item We first rigorously quantify the intrusion detectability and network coverage attainable by a bistatic radar network,
    while taking into account the complication that each pair of transmitter and receiver can potentially form a bistatic radar. We then study the worst-case coverage under deterministic deployment, aiming to find optimal deployment locations of radar transmitters and
    receivers such that the worst-case intrusion detectability is maximized. We present a sufficient condition on the field geometry under which it is optimal to deploy radars on a shortest line segment across the field, which we refer as a \emph{shortcut barrier}. The line barrier appears intuitive but does not hold for arbitrary field geometry, for which we give counter-examples. Further, when radars are deployed on the shortcut barrier, the worst-case intrusion detectability turns out to be the vulnerability of the shortcut barrier, which is the minimum detectability of all points on it.

\item A main thrust of this study is devoted to characterizing the optimal deployment locations of radars on the shortcut
    barrier to minimize its vulnerability. Since the optimization problem is neither differentiable nor convex, it is highly non-trivial to solve. We first quantify the local structure of detectability corresponding to a given deployment order and spacings along a shortcut barrier. Then, for a given deployment order, we establish the existence and optimality of balanced deployment spacings required to attain the minimum vulnerability. Next, we derive sufficient conditions for an optimal deployment order, and characterize the corresponding optimal deployment orders. \emph{Our findings reveal that the optimal deployment locations exhibit a balanced structure.}
    Furthermore, under the optimal deployment, it suffices for a receiver to form at most two bistatic radars to
    guarantee the optimal coverage quality, under the condition that the number of receivers is no less than that of
    transmitters (which is often the case since radar transmitters are more costly).



\item We also study the worst-case coverage under random deployment, and quantify the worst-case intrusion path for any
    given deployment of radars. In particular, by developing a novel \emph{2-site Voronoi diagram} with graph search techniques, we design an
    algorithm to find an approximate worst-case intrusion detectability, where the approximation error
    can be made arbitrarily small. The algorithm is shown to have polynomial-time complexity.

\end{itemize}

We believe that the studies we initiated here on bistatic radar networks scratch only the tip of the iceberg. There
are still many questions remaining open for the design of a networked radar system.

The rest of this paper is organized as follows. Section~\ref{sc:mdl} introduces the model of worst-case coverage for the
bistatic radar network. In Section~\ref{sc:dtm}, we characterize the optimal deployment of radars for maximizing the worst-case
detectability. Section~\ref{sc:rdm} presents an efficient algorithm for finding an approximate worst-case detectability given
arbitrary locations of radars. Section~\ref{sc:evl} provides the evaluation results. Related works are discussed in Section~\ref{sc:rlt}. The paper is concluded in Section~\ref{sc:ccl}.

\section{System Model}\label{sc:mdl}

\textit{Basic Setting.} We consider a \textit{bistatic radar network} deployed in a \textit{field of interest} $F$. The field
$F$ is a bounded and connected region enclosed by four curves: a left boundary $F_l$, a right boundary $F_r$, an entrance side, and a
destination side (see Figure~\ref{fg:barrier}). Note that we allow the boundary curves of the field to be arbitrary. An intruder (target) can traverse through $F$ along any \textit{intrusion path} $P\subset F$ from a point on the entrance to a point on the destination. Specifically, the bistatic radar network consists of $M$ radar transmitters $T_i \in \mathcal{T}$, $i \in \mathcal{M}\triangleq\{1,\cdots,M\}$ and $N$ radar receivers $R_j \in \mathcal{R}$, $j \in \mathcal{N}\triangleq \{1,\cdots,N\}$. We assume that each pair of transmitter and receiver can potentially form a
bistatic radar. We further assume that orthogonal transmissions are used for interference avoidance. For convenience, we also
use $T_i$ or $R_j$ to denote the location (point) of node $T_i$ or $R_j$, respectively.

\textit{Bistatic radar SNR and Cassini Oval.} For a bistatic radar, the received SNR for a target is given by
\begin{equation}\label{SNR}
{\mbox{SNR}} = \frac{K}{\|TX\|^2\|RX\|^2}
\end{equation}
where $\|TX\|$ and $\|RX\|$ denote the transmitter-target and receiver-target distances, respectively, and $K$ denote
\textit{bistatic radar constant} which reflects physical-layer system characteristics, such as transmitting power, radar cross
section, and transmitter/receiver antenna power gains. For ease of exposition, we assume the bistatic radar constant is the same, and our study on homogeneous bistatic radars here will serve as a major step for studying the heterogenous case. Given a bistatic radar, the SNR contours are characterized by the Cassini ovals with foci at the transmitter and receiver. We use $C_{T,R}(c)$ to denote the Cassini oval with foci at points $T$ and $R$ for constant distance product $c$.

\textit{Intrusion Detectability and Network Coverage.} Let $\|ab\|$ denote the (Euclidean) distance between two points $a$ and
$b$. It follows from \eqref{SNR} that the received SNR by a bistatic radar $\{T_i, R_j\}$ from a target at some point $p$ is
determined by the distance product $\|T_ip\|\|R_jp\|$. Then the \textit{detectability} of a point $p$ is quantified by the
minimum distance product from $p$ to all bistatic radars, denoted by
\begin{equation}\label{det}
I(p) \triangleq \min_{T_i\in\mathcal{T},R_j\in\mathcal{R}}\|T_ip\|\|R_jp\|.
\end{equation}
Observe from \eqref{det} that the detectability increases when $I(p)$ reduces, and vice versa. The detectability of an intruder is captured by the detectability of its intrusion path $P$, which is the maximum detectability of
all points on $P$, denoted by
\begin{equation}
B(P) \triangleq \min_{p\in P}I(p).
\end{equation}
From the radar network's perspective, in the worst case, the intruder traverses through $F$ along an intrusion path $P^*$ such
that $B(P^*)$ is maximized. We refer this path and its detectability as the \textit{worst-case intrusion path} and the
\textit{worst-case intrusion detectability} (WID), respectively.

\section{Deterministic Deployment}\label{sc:dtm}


We consider both deterministic deployment and random deployment scenarios, in the same spirit as the arbitrary/random network models in the seminal work by Gupta and Kumar \cite{Kumar00}. In the deterministic deployment case, our goal is to find optimal deployment locations of radar nodes (i.e., $M$ transmitters and $N$ receivers) in $F$ such that WID is maximized, i.e.,
\begin{align}\label{prb1}
& \underset{T_i \in F, R_j \in F}{\textrm{minimize}} \ \ B(P^*)
\end{align}
The above problem is highly non-trivial in general since the boundaries of field $F$ can be arbitrary. In particular, we are interested in the line-based deployment scheme in which radars are deployed along a line which is simple to implement in practice. Further, we will show that it is optimal under some mild condition on the field geometry. To the best of our knowledge, the optimality of line-based deployment for worst-case coverage has not been studied in literature.

\subsection{Shortcut Barrier-based Deployment}

We need the notion of barrier to establish our results on the line-based deployment. Given any $c > 0$, the \textit{detection
range} of a bistatic radar $\{T_i, R_j\}$, denoted by $A_{T_i, R_j}(c)$, is the region enclosed by the Cassini oval $C_{T_i,
R_j}(c)$ such that $\|T_ip\|\|R_jp\| \le c$ for all $p\in A_{T_i, R_j}(c)$. The detection range of the bistatic radar network,
denoted by $A(c)$, is the union of the detection ranges of all the bistatic radars. A \textit{barrier} is a curve in the
field $F$ connecting the left boundary $F_l$ and right boundary $F_r$ such that any intrusion path through $F$ intersects with
the barrier. The worst-case coverage is intimately related to the notion of barrier as given in the next property. The proof is based on an argument similar to the growing disks in \cite{Li03TC} and is omitted here.
\begin{pp}
The worst-case intrusion detectability \linebreak $B(P^*)$ is equal to the smallest value of $c$ such that there exists a barrier in the network's
detection range $A(c)$.
\end{pp}
We define the \textit{vulnerability} of a barrier $U$ as the minimum detectability of all points in $U$, denoted by
\begin{equation}\label{vul}
Q(U) \triangleq \max_{p\in U}I(p).
\end{equation}

\begin{figure}[t]
\centering
\includegraphics[width=0.45\textwidth]{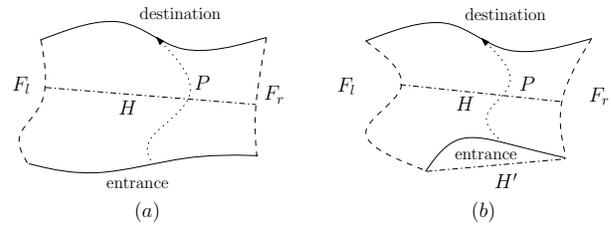}
\caption{(a) $H$ is a shortcut barrier; (b) $H$ is a shortest barrier but not a shortcut barrier since $H'$ is shorter than $H$; the optimal deployment may not be on $H$.} \label{fg:barrier}
\end{figure}

Intuitively, we expect to attain the best coverage quality by deploying radars along a shortest barrier which an intruder
must pass through. We make the following assumption on the geometry of field $F$.
\begin{as}\label{as:F}
There exists a shortest line segment $H$ connecting $F_l$ and $F_r$ such that $H \subset F$.
\end{as}
Although a shortest line segment connecting $F_l$ and $F_r$ always exists, it is critical to assume that it lies in the field $F$ (see Figure~\ref{fg:barrier}). We refer such an $H$ as the \textit{shortcut barrier} (SCB) and let $h$ denote its length. In other words, $h$ is the shortest distance between a point in $F_l$ and a point in $F_r$. Note that a shortcut barrier is always a shortest barrier that has the shortest length among all barriers, but the converse is not true (as illustrated in Figure~\ref{fg:barrier}). Assumption 1 captures a large class of field geometry, e.g., any $F$ of convex shape belongs to this class.

The next result shows that the SCB-based deployment is indeed optimal if the shortcut barrier exists. (All proofs of the theorems and lemmas in the sequel are relegated to Appendix).
\begin{thm}\label{thm:mbp}
Under Assumption 1, it is optimal to deploy radars on the shortcut barrier $H$, in order to maximize the worse-case
intrusion detectability.
\end{thm}

Although Theorem~\ref{thm:mbp} appears intuitive, we caution that it is non-trivial since the proof hinges on the existence of a shortcut barrier such that a line barrier with no greater vulnerability can be constructed from any arbitrary barrier. In other words, if the shortcut barrier does not exist, the optimal deployment may not be on a shortest barrier (see Figure~2(b)).

Let $H_l$ and $H_r$ denote the end points of $H$ with $H_l\in F_l$ and $H_r\in F_r$. Note that the worst-case intrusion detectability under a line-based deployment is no greater than, but not necessarily equal to, the vulnerability of the line segment. However, the equality holds for the SCB-based deployment as we show in the next result.
\begin{thm}\label{thm:bl}
Under Assumption 1, if radars are deployed on the shortcut barrier $H$, the worst-case intrusion detectability amounts to
the vulnerability of $H$, i.e., $B(P^*) = Q(H)$.
\end{thm}

Combining Theorem~\ref{thm:mbp}~and~\ref{thm:bl}, we conclude that the problem \eqref{prb1} is equivalent to finding the radar deployment on the shortcut barrier $H$ such that $Q(H)$ is minimized.

\subsection{Line-based Optimal Deployment Locations}

In this subsection, we study the optimal deployment locations of radars on the shortcut barrier $H$ to minimize $Q(H)$.
Let $t_i \triangleq \|H_lT_i\|$ and $r_j \triangleq \|H_lR_j\|$. Mathematically, our problem can be formulated as
\begin{align}\label{prb2}
&\underset{t_i,  r_j}{\textrm{minimize}} \quad \max_{0\le x\le h}\min_{i\in\mathcal{M},j\in\mathcal{N}} |x-t_i||x-r_j| \\
&\textrm{subject to} \quad  0\le t_i\le h, i\in\mathcal{M} \nonumber \\
& \qquad \qquad \quad \ 0\le r_j\le h, j\in\mathcal{N} \nonumber
\end{align}
where $\min_{i\in\mathcal{M},j\in\mathcal{N}}|x-t_i||x-r_j|$ represents the detectability of a point $p\in H$ with $\|H_lp\| =
x$. It can be easily checked that the objective function of problem~\eqref{prb2} is neither differentiable nor convex in general. Therefore, standard
optimization methods can not be applied here.

Consider the reformulation of problem~\eqref{prb2} as follows. First, we treat $H_l$ and $H_r$ as two virtual nodes and relax the constraint $\overline{H_lH_r} = h$. Suppose all nodes in $\mathcal{T}$ and $\mathcal{R}$ as well as $H_l$ and $H_r$ are deployed on a line such that $H_l$ and $H_r$ are the leftmost and rightmost nodes, respectively. Let $\mathbf{S} \triangleq (H_l,S_1,\cdots,S_J,H_r)$ denote a \textit{deployment order} (``order'' for short) of all nodes, where $J \triangleq M + N$ and $(S_1,\cdots,S_J)$ is a permutation of the set $\mathcal{T}\cup\mathcal{R}$ such that $\|H_lH_l\| = 0 \le \|H_lS_1\| \le \cdots \le \|H_lS_J\| \le \|H_lH_r\|$. Without ambiguity, $\mathbf{S}$ also denote an order of locations of all nodes. Let
$\mathbf{D}_{\mathbf{S}} = (\|H_lS_1\|,\cdots,\|S_JH_r\|)$ denote the \textit{deployment spacings} (``spacings'' for short) of
a given deployment order of nodes $\mathbf{S}$, which are the distances between all pairs of neighbor nodes in $\mathbf{S}$.
Similarly, let $(S_i,S_{i+1},\cdots,S_{i+j})$ denote a \textit{deployment suborder} (``suborder'' for short), which is a
deployment order of neighbor nodes in $\mathbf{S}$, and $\mathbf{D}_{(S_i,S_{i+1}\cdots,S_{i+j})} =
(\|S_iS_{i+1}\|,\cdots,\|S_{i+j-1}S_{i+j}\|)$ denote its deployment spacings. Clearly, any deployment order $\mathbf{S}$ with
any deployment spacings $\mathbf{D}_{\mathbf{S}}$ represent some deployment locations of $\mathcal{T}$ and $\mathcal{R}$ on a
line segment with length $\|H_lH_r\|$. Likewise, any deployment locations of $\mathcal{T}$ and $\mathcal{R}$ on the line
barrier $H$ can be represented by some deployment order $\mathbf{S}$ with some deployment spacings $\mathbf{D}_{\mathbf{S}}$
under the constraint $\|H_lH_r\| = h$. To summarize, the problem~\eqref{prb2} can be recast as
\begin{pb}
Find an optimal deployment order $\mathbf{S}^*$ with optimal deployment spacings $\mathbf{D}^*_{\mathbf{S}^*}$ under the
constraint $\|H_lH_r\| = h$ such that $Q(\overline{H_lH_r})$ is minimized.
\end{pb}

In what follows, we outline the main steps to solve Problem~1.
\begin{enumerate}
\item[Step1:] We show a general and important structure of detectability on $\overline{H_lH_r}$ (Lemma~\ref{lm:local}), which leads to the concept of balanced spacings.

\item[Step2:] We introduce the concept of local suborder. Then we show the existence and characterize balanced spacings for a local suborder (Lemma~\ref{lm:bal_loc}), and establish the optimality of balanced spacings for a local suborder (Lemma~\ref{lm:bal_loc_opt}).

\item[Step3:] We rule out some orders from consideration, which leads to the concept of candidate order (Lemma~\ref{lm:swap}). Then we
    show that a candidate order consists of decoupled local suborders, such that the results in Step 2 can be applied to show that balanced spacings exist and are optimal for a candidate order (Theorem~\ref{thm:bal_opt}).

\item[Step4:] Based on the optimal spacings obtained above, we derive sufficient conditions for an optimal order (Theorem~\ref{thm:ord_opt}). Then we characterize the optimal orders.
\end{enumerate}


We start with two observations. First, swapping the locations of any pair of transmitters or any pair of receivers results in an equivalent deployment. Second, transmitters and receivers are reciprocal to each other. Specifically, replacing all transmitters by receivers and replacing all receivers by transmitters results in an equivalent deployment. These observations will be used repeatedly in the sequel.

\begin{figure}[t]
\centering
\includegraphics[width=0.45\textwidth]{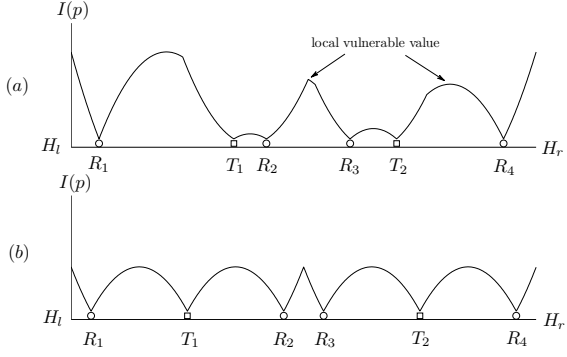}
\caption{The detectability of a point attains local minimums at the end points and the midpoints of all pairs of neighbor radar nodes.
The local vulnerable values are (a) unequal with arbitrary deployment spacings; and (b) equal with balanced deployment spacings.} \label{fg:structure}
\end{figure}

Let $Y_{ab}$ denote the midpoint between two points $a$ and $b$. In the next lemma, we show a local structure
for the vulnerability of $\overline{H_lH_r}$.
\begin{lm}\label{lm:local}
Given any order $\mathbf{S}$ with any spacings $\mathbf{D}_{\mathbf{S}}$, $I(p)$ attains local maximums on $\overline{H_lH_r}$
at the end nodes $H_l$ and $H_r$, and at the midpoints of all pairs of neighbor nodes in $(S_1,\cdots,S_J)$ (as depicted in
Figure~\ref{fg:structure}). In particular,
\begin{align*}
&\quad  \arg\max_{p\in \overline{H_lS_1}}I(p) = H_l, \quad \arg\max_{p\in \overline{S_JH_r}}I(p) = H_r \\
& \arg\max_{p\in \overline{S_iS_{i+1}}}I(p) = Y_{S_iS_{i+1}}, \ \ \forall i\in\{1,\cdots,J-1\}.
\end{align*}
\end{lm}
For convenience, we refer a local maximum point of $I(p)$ and its value as a \textit{local vulnerable point} and \textit{local
vulnerable value}, respectively. The local structure presented in Lemma~\ref{lm:local} is essentially due to that the detectability of a point is dominated by the closet transmitter and receiver. Therefore, it suffices to characterize the local vulnerable values at the
local vulnerable points to determine the vulnerability of $\overline{H_lH_r}$. We say the spacings of an order of nodes are
\textit{balanced} if all the local vulnerable values between or on the two end nodes of that order are equal (see
Figure~\ref{fg:structure}).

We define the \textit{pattern} of an order of nodes as the order of node types (transmitter type $T$ or receiver type $R$) of
those nodes. We use $T^k$ or $R^k$ to denote $k$ consecutive $T$ or $R$ in a pattern, respectively. We define three kinds of
\textit{local patterns} $\mathbf{P}_1 = (T, R^k, H_r)$, $(R, T^k, H_r)$, $(H_l, R^k, T)$, or $(H_l, T^k, R)$ for $k\ge1$,
$\mathbf{P}_2$ $=$ $(T, R^k, T)$ or $(R, T^k, R)$ for $k\ge2$, and $\mathbf{P}_3 = (T, R)$ or $(R, T)$. We say a suborder
$\mathbf{S}_i$ is a \textit{local suborder} if it has a local pattern, and it is associated with a \textit{local zone}, denoted
by $Z_{\mathbf{S}_i}$, which is the line segment between the two end nodes of that local suborder. For example, if
$\mathbf{S}_i = (T_1,R_1,\cdots,R_k,H_r)$, then $Z_{\mathbf{S}_i} = \overline{T_1H_r}$. The vulnerability of a local zone
$Z_{\mathbf{S}_i}$ is $Q(Z_{\mathbf{S}_i})$. We denote the length of $Z_{\mathbf{S}_i}$ by $L_{\mathbf{S}_i}$.

We observe that for any local suborder $\mathbf{S}_i$ with any spacings, the closest transmitter and receiver from $\mathbf{S}$
for any local vulnerable point in $Z_{\mathbf{S}_i}$ are nodes from $\mathbf{S}_i$. For example, if $\mathbf{S}_i = (T_1,R_1)$,
we can see that the closest transmitter and receiver for $Y_{T_1R_1}$ are $T_1$ and $R_1$, respectively; if $\mathbf{S}_i =
(T_1,R_1,\cdots,R_k,H_r)$, the closest transmitter for any of $Y_{T_1R_1}$, $\cdots$, $Y_{R_{k-1}R_k}$, $H_r$, is $T_1$, and
the closest receiver for any of $Y_{T_1R_1}$, $\cdots$, $Y_{R_{k-1}R_k}$, $H_r$, is from $\{R_1,\cdots,R_k\}$. Hence, the
vulnerability of a local zone $Z_{\mathbf{S}_i}$, i.e., $Q(Z_{\mathbf{S}_i})$, only depends on the locations of nodes in
$\mathbf{S}_i$, or in other words, depends on the spacings $\mathbf{D}_{\mathbf{S}_i}$, and it is independent of the locations
of nodes not in $\mathbf{S}_i$. Similar observations can be made for any local suborder. Using Lemma~\eqref{lm:local}, in the
next result, we characterize the balanced spacings of a local suborder.

\begin{lm}\label{lm:bal_loc}
For any $c > 0$, let $e_0(c) \triangleq 2\sqrt{c}$ and $e_j(c)$ be the unique positive value of $x$ such that
$(\sum^{j-1}_{i=0}e_i(c) + x/2)(x/2) = c$ for any $j \ge 1$. Given any $c > 0$ and any local suborder $\mathbf{S}_i$, there
exist unique spacings $\mathbf{D}_{\mathbf{S}_i}$ such that $\mathbf{D}_{\mathbf{S}_i}$ is balanced with $Q(Z_{\mathbf{S}_i}) =
c$. Furthermore, it is given by, e.g., if $\mathbf{S}_i = (T_1,R_1)$, then $\mathbf{D}_{\mathbf{S}_i} = (e_0)$; \\if
$\mathbf{S}_i = (T_1,R_1,\cdots$, $R_k,H_r)$, then
\[\mathbf{D}_{\mathbf{S}_i} = (e_0, e_1, \cdots, e_{k-1}, e_k/2);\]
if $\mathbf{S}_i = (T_1,R_1,\cdots,R_k,T_2)$ and $k$ is even, then
\[\mathbf{D}_{\mathbf{S}_i} = (e_0,e_1,\cdots, e_{k/2-1}, e_{k/2}, e_{k/2-1}, \cdots, e_1, e_0)\]
or if $k$ is odd, then
\[\mathbf{D}_{\mathbf{S}_i} = (e_0,e_1,\cdots, e_{(k-1)/2}, e_{(k-1)/2}, \cdots, e_1, e_0)\]
where the arguments ``$(c)$'' of $e_i(c)$ are omitted for brevity. Similar results can be obtained for any other local
suborder.
\end{lm}
By definition, given $c$, the value of $e_i(c)$, $i\ge0$ can be found iteratively and it decreases as $i$ increases (see Table~1).

\begin{table}\label{tbl}
\caption{Values of balanced spacings}
\centering
\begin{tabular}{ | l | l | l | l | l | l |}
    \hline
    c & $e_0(c)$ & $e_1(c)$ & $e_2(c)$ & $e_3(c)$ & $e_4(c)$  \\ \hline
    1  & 2.0000  &  0.8284  &  0.6357  &  0.5359  &  0.4721     \\
    5  & 4.4721  &  1.8524  &  1.4214  &  1.1983  &  1.0557     \\
    10 & 6.3246  &  2.6197  &  2.0102  &  1.6947  &  1.4930     \\
    20 & 8.9443  &  3.7048  &  2.8428  &  2.3966  &  2.1115     \\ \hline
\end{tabular}
\end{table}

The following lemma shows that balanced spacings are optimal for a local suborder.
\begin{lm}\label{lm:bal_loc_opt}
Given any local suborder $\mathbf{S}_i$, suppose $\mathbf{D}_{\mathbf{S}_i}$ is balanced with $Q(Z_{\mathbf{S}_i}) = c$. If
$\mathbf{S}'_i$ is the same local suborder as $\mathbf{S}_i$ and with spacings $\mathbf{D}_{\mathbf{S}'_i}$ such that
$Q(Z_{\mathbf{S}'_i}) \le c$, then $L_{\mathbf{S}'_i} \le L_{\mathbf{S}_i}$. Furthermore, if $L_{\mathbf{S}'_i} =
L_{\mathbf{S}_i}$, then $\mathbf{D}_{\mathbf{S}'_i} = \mathbf{D}_{\mathbf{S}_i}$.
\end{lm}
Lemma~\ref{lm:bal_loc_opt} suggests that given a local suborder and a constraint on the local zone's vulnerability, the balanced spacings maximize the length of the local zone.

We need the following lemma which reduces our search space significantly.
\begin{lm}\label{lm:swap}
There exists an optimal order $\mathbf{S}$ \textbf{not} having a suborder with any of the following patterns: $(T,T$, $R,R)$,
$(R,R,T,T)$, $(T,T,R,H_r)$, $(R,R,T,H_r)$, $(H_l,T$, $R,R)$, and $(H_l,R,T,T,)$.
\end{lm}
We say an order $\mathbf{S}$ is a \textit{candidate} order if it does not have a suborder with any of the patterns given in
Lemma~\ref{lm:swap}.

We observe that any given candidate order $\mathbf{S}$ consists of a series of local suborders
$\mathbf{S}_1,\cdots,\mathbf{S}_m$ such that 1) each node in $\mathbf{S}$ is included in some $\mathbf{S}_i$; 2) the last node
of $\mathbf{S}_i$ is the first node of $\mathbf{S}_{i+1}$ for all $i = 1,\cdots,m-1$. In particular, $\mathbf{S}_1$ and
$\mathbf{S}_m$ have patterns $\mathbf{P}_1$, and $\mathbf{S}_i$ has pattern $\mathbf{P}_2$ or $\mathbf{P}_3$ for $i =
2,\cdots,m-1$. For example,
\begin{align*}
\mathbf{S} = (\lefteqn{\overbrace{\phantom{H_l,R_1,R_2,T_1}}^{\mathbf{S}_1}}H_l,R_1,R_2,\lefteqn{\underbrace{\phantom{T_1,R_3,R_4,R_5,T_2}}_{\mathbf{S}_2}}
T_1,R_3,R_4,R_5,\lefteqn{\overbrace{\phantom{T_2,R_6}}^{\mathbf{S}_3}}T_2,\underbrace{R_6,T_3,T_4,H_r}_{\mathbf{S}_4}).
\end{align*}
Accordingly, $\overline{H_lH_r}$ consists of a series of local zones $Z_{\mathbf{S}_1}$, $\cdots$, $Z_{\mathbf{S}_m}$, and
$\mathbf{D}_{\mathbf{S}}$ consists of a series of spacings $\mathbf{D}_{\mathbf{S}_1}$, $\cdots$, $\mathbf{D}_{\mathbf{S}_m}$.
To see why such a series of local suborders exists for any given candidate order $\mathbf{S}$, we construct a \textit{super
order} $\mathbf{S}^+$ from $\mathbf{S}$ by combining neighbor nodes of the same type in $\mathbf{S}$ into a \textit{super
node}. A node not combined in $\mathbf{S}$ is a \textit{simple node} in $\mathbf{S}^+$. For the above example, the super order
is given by
\begin{align*}
\mathbf{S}^+ = (H_l,R^+_{1,2},T_1,R^+_{3,4,5},T_2,R_6,T^+_{3,4},H_r).
\end{align*}
It is clear by our construction that two neighbor nodes in $\mathbf{S}^+$ (excluding $H_l$ and $H_r$) are of different types
(transmitter type or receiver type). Since $\mathbf{S}$ is a candidate order, it does not have a suborder with any pattern
given in Lemma~\ref{lm:swap}. Hence, it can be easily checked that two neighbor nodes in $\mathbf{S}^+$ can not be both super
nodes, and a super node can not be the third or third-to-last node in $\mathbf{S}^+$. Then we can see that the first three
nodes and the last three nodes in $\mathbf{S}^+$ represent two local suborders with patterns $\mathbf{P}_1$; each super node
and its two neighbor simple nodes in $\mathbf{S}^+$ represent a local suborder with pattern $\mathbf{P}_2$; two neighbor simple
nodes not at the beginning or end of $\mathbf{S}^+$ (excluding $H_l$ and $H_r$) represent a local suborder with pattern
$\mathbf{P}_3$.



The decoupled structure of a candidate order allows us to apply results of local suborders. Consider a given candidate order
$\mathbf{S}$. We know that $\mathbf{S}$ consists of a series of local suborders $\mathbf{S}_1,\cdots,\mathbf{S}_m$. As a
result, the set of local vulnerability values on $\overline{H_lH_r}$ is a union of disjoint sets of local vulnerability values
on all the local zones $Z_{\mathbf{S}_1}$, $\cdots$, $Z_{\mathbf{S}_m}$. Therefore, since local vulnerability values on a local zone $Z_{\mathbf{S}_i}$ only depend on the corresponding spacings $\mathbf{D}_{\mathbf{S}_i}$, we can see that
$\mathbf{D}_{\mathbf{S}}$ is balanced with $Q(\overline{H_lH_r}) = c$ if and only if $\mathbf{D}_{\mathbf{S}_i}$ is balanced
with $Q(Z_{\mathbf{S}_i}) = c$ for all $i= 1,\cdots,m$. Hence, by Lemma~\ref{lm:bal_loc}, there exists unique
$\mathbf{D}_{\mathbf{S}}$ such that $\mathbf{D}_{\mathbf{S}}$ is balanced with $Q(\overline{H_lH_r}) = c$.

We observe that given a candidate order $\mathbf{S}$, under the constraint that $\mathbf{D}_{\mathbf{S}}$ is balanced with
$Q(\overline{H_lH_r}) = c$, $\mathbf{D}_{\mathbf{S}}$ varies as a function of $c$. It follows from Lemma~\ref{lm:bal_loc} that
each of $\|H_lS_1\|$, $\cdots$, $\|S_JH_r\|$ is an increasing function of $c$, and hence $\|H_lH_r\| = \|H_lS_1\| + \cdots +
\|S_JH_r\|$ is also an increasing function of $c$. Furthermore, it can be easily verified that $\|H_lH_r\| = 0$ when $c = 0$
and $\|H_lH_r\|\rightarrow \infty$ when $c \rightarrow \infty$. Thus, there exists a unique $c > 0$ such that $\|H_lH_r\| = h$.
This implies that under the constraint $\|H_lH_r\| = h$, there exist unique balanced $\mathbf{D}_{\mathbf{S}}$. We summary this
result below as a corollary of Lemma~\ref{lm:bal_loc}.
\begin{cor}\label{cr:bal_unq}
Given any candidate order $\mathbf{S}$, under the constraint $\|H_lH_r\| = h$, there exist unique spacings
$\mathbf{D}_{\mathbf{S}}$ such that $\mathbf{D}_{\mathbf{S}}$ is balanced.
\end{cor}

Applying Lemma~\ref{lm:bal_loc_opt}, we obtain the following theorem.
\begin{thm}\label{thm:bal_opt}
Given any candidate order $\mathbf{S}$,  the unique balanced spacings
$\mathbf{D}_{\mathbf{S}}$ minimize $Q(\overline{H_lH_r}) $ under the constraint $\|H_lH_r\| = h$.
\end{thm}
Theorem~\ref{thm:bal_opt} suggests that the optimal spacings for any given candidate order is the unique balanced spacings
referred in Corollary~\ref{cr:bal_unq}.

Since we have found the optimal spacings $\mathbf{D}^*_{\mathbf{S}}$ for any candidate order $\mathbf{S}$, our next step is to
find an optimal order $\mathbf{S}^*$. Suppose the number of transmitters is no greater than the number of receivers, i.e., $M
\le N$, WLOG. Since all transmitters are equivalent, we suppose that transmitters in $\mathbf{S}$ are indexed by their order in
$\mathbf{S}$ such that $0\le\|H_lT_1\|\le\cdots\le\|H_lT_M\|\le\|H_lH_r\|$. Define $\mathbf{N} \triangleq
(n_{H_lT_1},n_{T_1T_2},\cdots,n_{T_{M-1}T_M},n_{T_MH_r})$ where $n_{T_iT_{i+1}}$, $n_{H_lT_1}$, and $n_{T_MH_r}$ denote the
number of receivers between $T_i$ and $T_{i+1}$, between $H_l$ and $T_1$, and between $T_M$ and $H_r$, respectively, in
$\mathbf{S}$.  Since all receivers are equivalent, it suffices to determine $n_{T_iT_{i+1}}$, $n_{H_lT_1}$, and
$n_{T_MH_r}$ for an optimal order. The next result provides sufficient conditions for the optimality of an order.
\begin{thm}\label{thm:ord_opt}
A candidate order $\mathbf{S}$ is optimal if
\begin{align*}
&|n_{T_iT_{i+1}} - n_{T_jT_{j+1}}| \le 1, \forall i\neq j \\
&|n_{T_iT_{i+1}} - 2n_{H_lT_1}| \le 1, |n_{T_iT_{i+1}} - 2n_{T_MH_r}| \le 1, \forall i.
\end{align*}
\end{thm}

Using the optimality conditions in Theorem~\ref{thm:ord_opt}, we can find an optimal order $\mathbf{S}^*$ described as
follows. Let two integers $q$ and $r$ be the quotient and remainder of $N/M$. If $q$ is even, there exists an $\mathbf{S}^*$
such that
\[\mathbf{N}^* = (\frac{q}{2},\overbrace{q+1,\cdots,q+1}^{r},\overbrace{q,\cdots,q}^{M-1-r},\frac{q}{2});\]
if $q$ is odd and $r = 0$, there exists an $\mathbf{S}^*$ such that
\[\mathbf{N}^* = (\frac{q+1}{2},\overbrace{q,\cdots,q}^{M-1},\frac{q-1}{2});\]
if $q$ is odd and $r \ge 1$, there exists an $\mathbf{S}^*$ such that
\[\mathbf{N}^* = (\frac{q+1}{2},\overbrace{q+1,\cdots,q+1}^{r-1},\overbrace{q,\cdots,q}^{M-r},\frac{q+1}{2}).\]
It can be easily seen that any order obtained from the above optimal order by exchanging the values of $n_{H_lT_1}$ and
$n_{T_MH_r}$, or the values of $n_{T_iT_{i+1}}$ and $n_{T_jT_{j+1}}$ for some $i\neq j$, also satisfies the optimality
condition, and hence is optimal. We observe that the optimal order also exhibits a balanced structure.

\begin{figure}[t]
\vspace{-1cm}
\centering
\includegraphics[width=0.50\textwidth]{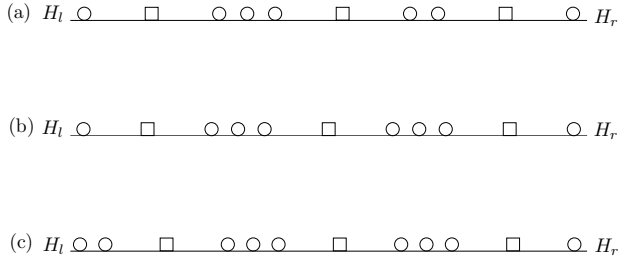}
\caption{Optimal deployment locations of radar transmitters (square) and receivers (circle) for (a) $M = 3$ and $N = 7$ (b) $M = 3$ and $N = 8$
(c) $M = 3$ and $N = 9$.} \label{fg:exp}
\end{figure}

Given an optimal order $\mathbf{S}^*$, by Theorem~\ref{thm:bal_opt}, the optimal spacings $\mathbf{D}^*_{\mathbf{S}^*}$ is the
unique balanced spacings under the constraint $\|H_lH_r\| = h$. The optimal spacings can be found by a bisection search. In
each search step, for a search criterion $c$, we can obtain by Lemma~\ref{lm:bal_loc} the unique balanced spacings
$\mathbf{D}_{\mathbf{S}^*}$ with $Q(\overline{H_lH_r}) = c$, and hence obtain $\|H_lH_r\|$. If $\|H_lH_r\| > h$, $c$ is
decreased for the next step; if $\|H_lH_r\| < h$, $c$ is increased for the next step. The search completes if $\|H_lH_r\| = h$.
Upon completion, the optimal spacings $\mathbf{D}^*_{\mathbf{S}^*}$ is found and the optimal value of $Q(\overline{H_lH_r})$ is
equal to the search criterion $c$.


\section{Random Deployment}\label{sc:rdm}

Next, we consider random deployment that can also be of great interest. In this scenario, given arbitrary locations of radars resulted from random deployment, we aim to find the worst-case intrusion path $P^*$ and the worst-case intrusion detectability $B(P^*)$. Due to the complex geometry of bistatic radar SNR, it is difficult to find $P^*$ accurately. Therefore, we design an efficient algorithm for finding an intrusion path $P'$ whose detectability is arbitrarily close to $B(P^*)$. Our algorithm partitions the field $F$ into sub-regions based on a novel 2-site Voronoi diagram, and then constructs a weighted graph $G$ for the sub-regions to search for an approximate worst-case intrusion path $P'_G$.

\begin{figure}[t]
\centering
\includegraphics[width=0.40\textwidth]{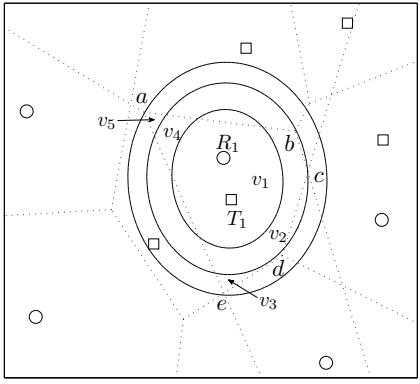}
\vspace{-1cm}
\caption{Overlapping Voronoi diagrams for radar transmitters and receivers.
The polygon with indices $\{a,b,c,d,e\}$ is a 2-site Voronoi region with respect to $T_1$ and $R_1$;
it is divided into sub-regions $\{v_1,v_2,v_3,v_4,v_5\}$.} \label{fg:2voronoi}
\end{figure}

Given a set of points $S$, the \textit{Voronoi region} $V_S(s)$ of a point $s\in S$ is the set of points which are closer to $s$ than to any other point in $S$, i.e.,
\begin{equation}\label{vor}
V_S(s) \triangleq \{p|\min_{s'\in S} \|ps'\| = \|ps\|\}.
\end{equation}
The Voronoi diagram for $S$ is the collection of all Voronoi regions $\{V_S(s), s\in S\}$. To partition the field $F$, we first construct Voronoi diagrams for the set of transmitters $\mathcal{T}$ and the set of receivers $\mathcal{R}$, respectively, i.e., $\{V_{\mathcal{T}}(T_i), T_i\in\mathcal{T}\}$ and $\{V_{\mathcal{R}}(R_j), R_j\in\mathcal{R}\}$, within the boundaries of $F$. Then $F$ can be partitioned into \textit{2-site Voronoi regions}, where a 2-site Voronoi region, denoted by $V_{\mathcal{T},\mathcal{R}}(T_i, R_j)$, is the intersection of a transmitter Voronoi region and a receiver Voronoi region, i.e.,
\begin{equation}\label{2vor}
V_{\mathcal{T},\mathcal{R}}(T_i, R_j) \triangleq V_{\mathcal{T}}(T_i) \cap V_{\mathcal{R}}(R_j)
\end{equation}
for all $T_i\in \mathcal{T}$ and $R_j\in \mathcal{R}$ such that $V_{\mathcal{T}}(T_i) \cap V_{\mathcal{R}}(R_j) \neq
\varnothing$ (see Figure~\ref{fg:2voronoi}). Choose an arbitrary $\epsilon > 0$. For each 2-site Voronoi region
$V_{\mathcal{T},\mathcal{R}}(T_i, R_j)$, we plot a sequence of Cassini ovals $C_{T_i, R_j}(k\epsilon)$ for consecutive positive
integer values of $k$ such that $V_{\mathcal{T},\mathcal{R}}(T_i, R_j)$ is further divided by these Cassini ovals into
sub-regions where each sub-region is bounded between two Cassini ovals with distance products being consecutive multiples of
$\epsilon$.

We construct a graph $G = (V, E)$ where each vertex represents a sub-region, and two vertices are connected by an edge if they
are adjacent sub-regions. Then we assign each vertex $v_i$ a weight $w_i$, which is equal to the smaller distance product of
the two Cassini ovals bounding this sub-region. We add two virtual vertices $s$ and $t$ to represent the entrance and
destination of $F$, respectively. There exists an edge between vertex $s$ and $v_i$ for $v_i\in V\setminus\{s,t\}$ if
sub-region $v_i$ is adjacent to the entrance. Similarly, we add an edge between vertex $t$ and $v_i$ for $v_i\in
V\setminus\{s,t\}$ if sub-region $v_i$ is adjacent to the destination.

It is clear that any intrusion path $P$ is represented by an $s-t$ path in $G$, denoted by $P_G$. We define the weight of an
$s-t$ path $P_G$, denoted by $W(P_G)$, as $W(P_G) \triangleq \min_{\{i:v_i\in P_G\}}w_i$. We aim to find an $s-t$ path $P'_G$
with maximum weight. This can be obtained by a bisection search between the smallest and largest vertex weights in $G$. In each
step, breadth-first-search is used to check the existence of a $s-t$ path using only vertices with weights larger than a search
criterion $c$. If a path exists, $c$ is increased to restrict the vertices considered in the next search step; otherwise, $c$ is
decreased to relax the constraint on the search. Upon completion, a $s-t$ path with maximum weight is found.

Suppose $M = O(n)$ and $N = O(n)$. The number of vertices in graph $G$, i.e., the number of sub-regions in field $F$, is $O(n^8/\epsilon^2 + n^6/\epsilon + n^4)$ and the reason is as follows. We can treat the field $F$ as a planar graph, where the vertices are the crossing points of Voronoi edges and Cassini ovals, and the edges are line segments or curves between two crossing points. The number of Voronoi edges and the number of Voronoi vertices are both $O(n)$. So the number of crossing points of Voronoi edges is $O(n^2)$. Hence, the number of 2-site Voronoi vertices is $O(n^2)$, and the number of 2-site Voronoi edges is $O(n^4)$. Since the field $F$ is bounded, the number of Cassini ovals plotted for a 2-site Voronoi region is $O(1/\epsilon)$. So the total number of crossing points of Cassini ovals with 2-site Voronoi edges is $O(n^4/\epsilon)$. Hence, the number of vertices of sub-regions is $O(n^4/\epsilon+n^2)$, and the number of edges of sub-regions is $O(n^8/\epsilon^2 + n^6/\epsilon + n^4)$. By Euler's formula, the number of faces, which is the number of sub-regions, is 2 - $O(n^4/\epsilon+n^2)$ + $O(n^8/\epsilon^2 + n^6/\epsilon + n^4)$, equal to $O(n^8/\epsilon^2 + n^6/\epsilon + n^4)$.

In the next result, part (a) follows from the definition of 2-site Voronoi diagram and our graph construction; part (b) follows from our result on the number of vertices in graph $G$. The proof is omitted due to space limitation.

\begin{thm}\label{thm:apx}
(a) The maximum path weight obtained by our proposed approximation algorithm is within $\epsilon$ to the worst-case detectability, i.e.,
\begin{equation}
B(P^*) - \epsilon \le W(P'_G) \le B(P^*).
\end{equation}
(b) Given $\epsilon$, our proposed approximation algorithm has polynomial-time complexity.
\end{thm}


\section{Performance Evaluation}\label{sc:evl}

In this section, we present some numerical results to illustrate the effectiveness of our proposed optimal radar deployment scheme.

\begin{figure}[t]
\vspace{-1cm}
\centering
\includegraphics[width=0.50\textwidth]{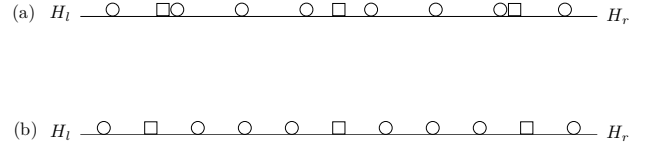}
\caption{Heuristic deployment locations of radar transmitters (square) and receivers (circle) for $M = 3$ and $N = 8$ (a) HEU-1 (b) HEU-2.} \label{fg:exp2}
\end{figure}

We examine the vulnerability of a line segment $H$ under the line-based deployment. In particular, we compare the optimal deployment scheme (OPT) with two heuristic deployment schemes. The first heuristic (HEU-1) is to deploy radar transmitters (or receivers, respectively) with uniform spacings such that the maximum distance from a point on $H$ to its closet transmitter (or receiver, respectively) is minimized (see Figure~\ref{fg:exp2}). Specifically, HEU-1 results in $2\|H_lT_1\| = \|T_1T_2\| \linebreak = \cdots = \|T_{M-1}T_M\| = 2\|T_MH_r\|$ and $2\|H_lR_1\| = \linebreak \|R_1R_2\| = \cdots = \|R_{N-1}R_N\| = 2\|R_NH_r\|$, where transmitters and receivers are indexed by their orders in $\mathbf{S}$, respectively. It can be easily verified that this scheme is optimal if we treat radar transmitters (or receivers, respectively) as traditional sensors with disk-based sensing model. The second heuristic (HEU-2) is to deploy radar transmitters and receivers according to an optimal order $\mathbf{S}^*$ from OPT but with uniform spacings such that the maximum distance from a point on $H$ to its closet transmitter or receiver is minimized (see Figure~\ref{fg:exp2}). Specifically, HEU-2 results in $2\|H_lS^*_1\| = \|S^*_1S^*_2\| = \cdots = \|S^*_{M-1}S^*_M\| = 2\|S^*_MH_r\|$.

\begin{figure}[h]
\centering
\includegraphics[width=0.25\textwidth]{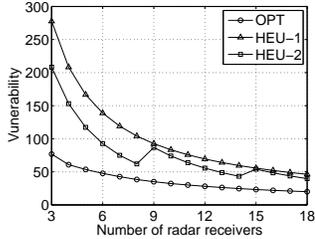}
\vspace{-0.5cm}
\caption{Vulnerability vs. number of radar receivers for 3 radar transmitters.}
\label{fg:sim1}
\end{figure}

\begin{figure}[h]
\centering
\includegraphics[width=0.25\textwidth]{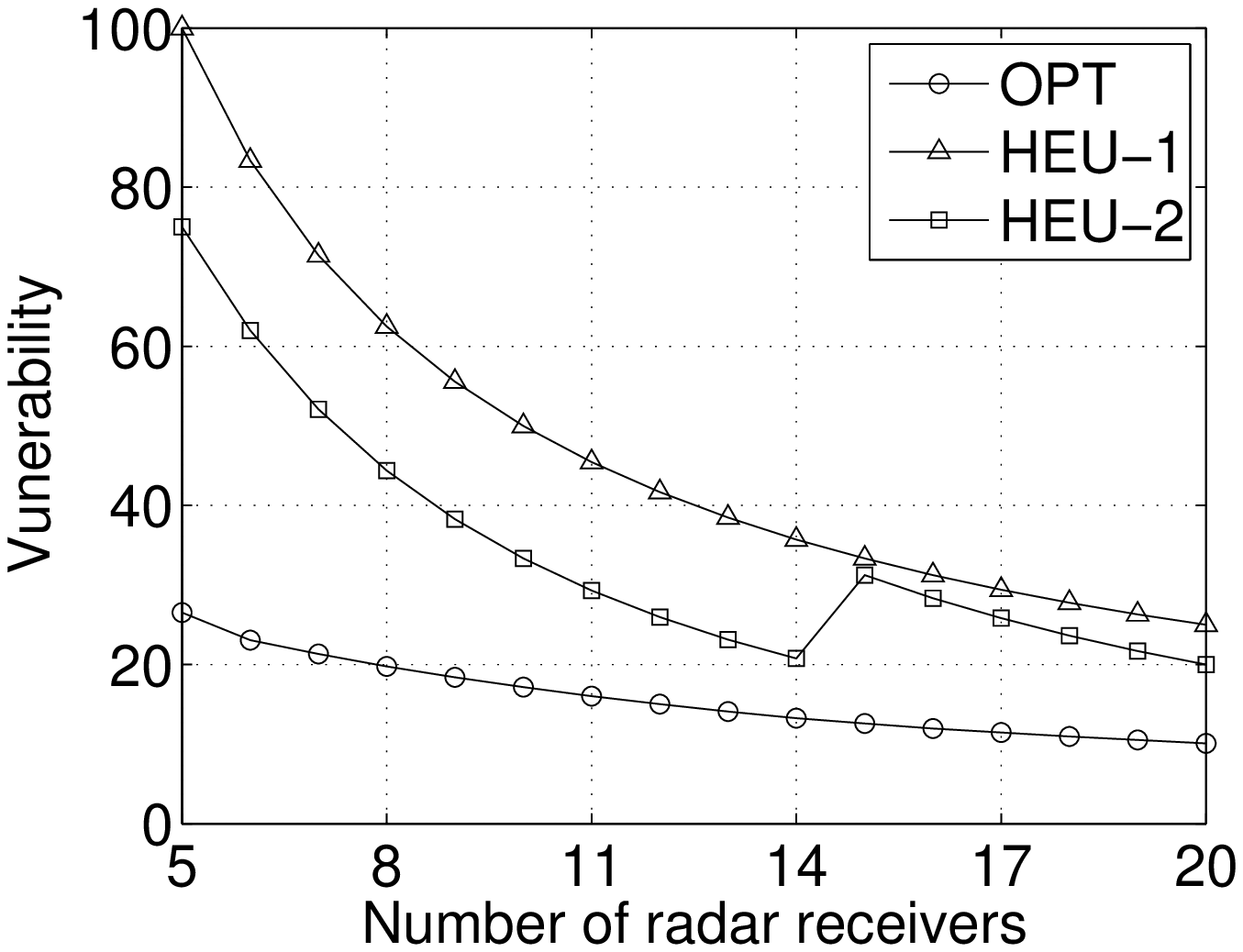}
\vspace{-0.5cm}
\caption{Vulnerability vs. number of radar receivers for 5 radar transmitters.}
\label{fg:sim2}
\end{figure}

\begin{figure}[h]
\centering
\includegraphics[width=0.25\textwidth]{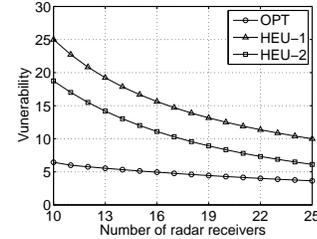}
\vspace{-0.5cm}
\caption{Vulnerability vs. number of radar receivers for 10 radar transmitters.}
\label{fg:sim3}
\end{figure}

Figure~\ref{fg:sim1}, Figure~\ref{fg:sim2} and Figure~\ref{fg:sim3} depict the vulnerability of $H$ versus the number of radar receivers for $3$, $5$, and $10$ radar transmitters, respectively, where we set $h = 100$. We note that HEU-2 results in considerably lower vulnerability than HEU-1. This is because HEU-1 deploys transmitters and receivers independently, while HEU-2 takes into account the joint design of transmitters and receivers. We observe that OPT further outperforms HEU-2 significantly. This suggests that optimal spacings is critical for improving the performance due to the complex geometry of bistatic radar SNR.


Next, we evaluate the worst-case intrusion detectability under random deployment using our proposed approximate algorithm. We consider a square field $F$ of size $100\times100$, where two opposite sides are entrance and destination, respectively, and the other two opposite sides are left and right boundaries, respectively. We consider deploying transmitters and receivers in the field randomly (RAN) with uniform distribution.

\begin{figure}[h]
\centering
\includegraphics[width=0.25\textwidth]{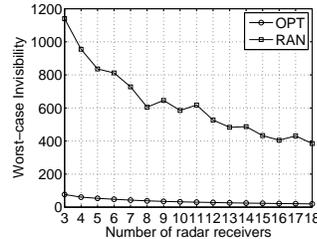}
\vspace{-0.5cm}
\caption{Worst-case detectability vs. number of radar receivers for 3 radar transmitters.}
\label{fg:sim4}
\end{figure}

\begin{figure}[h]
\centering
\includegraphics[width=0.25\textwidth]{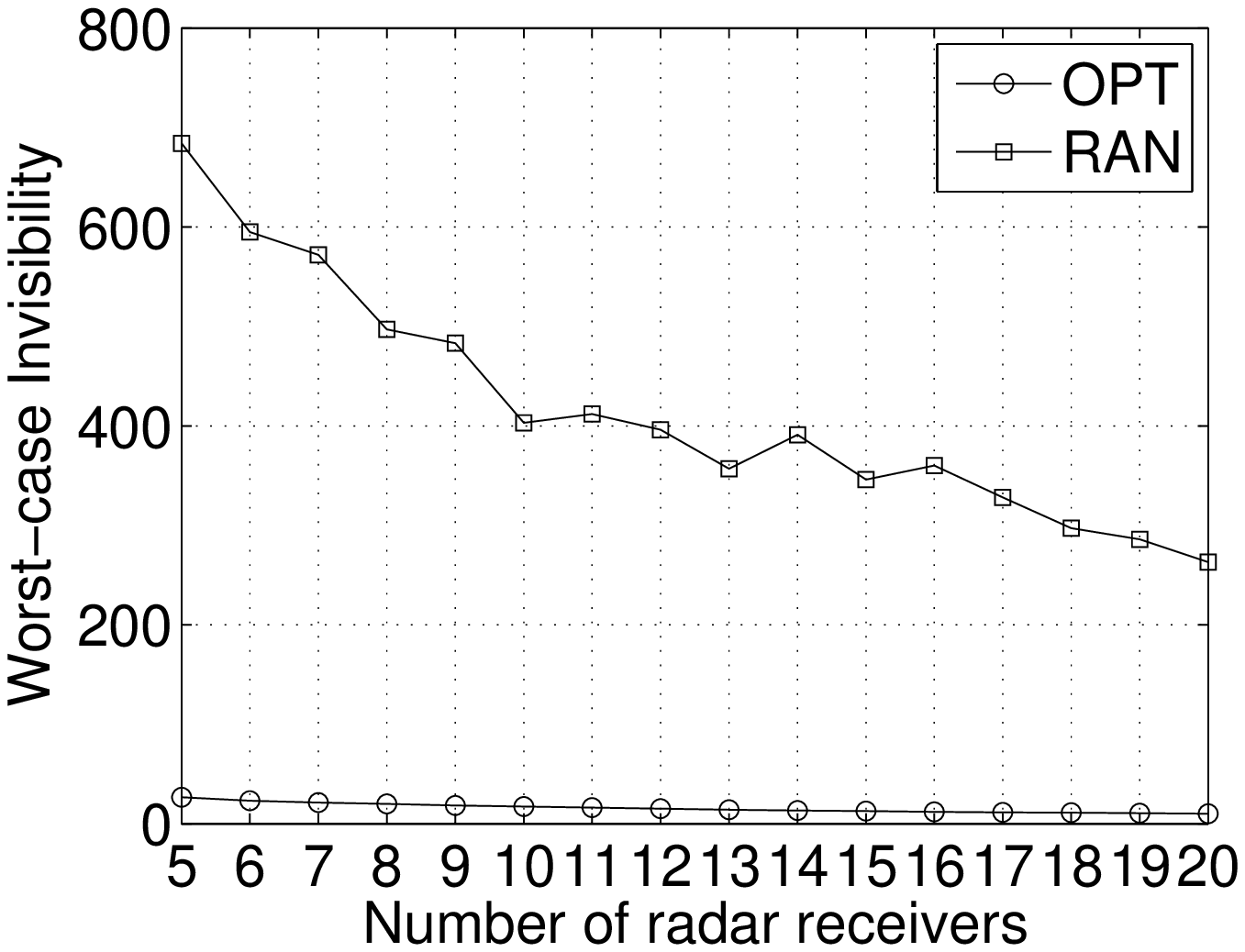}
\vspace{-0.5cm}
\caption{Worst-case detectability vs. number of radar receivers for 5 radar transmitters.}
\label{fg:sim5}
\end{figure}

\begin{figure}[h]
\centering
\includegraphics[width=0.25\textwidth]{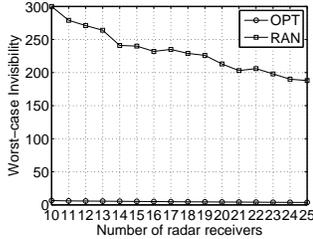}
\vspace{-0.5cm}
\caption{Worst-case detectability vs. number of radar receivers for 10 radar transmitters.}
\label{fg:sim6}
\end{figure}

Figure~\ref{fg:sim4}, Figure~\ref{fg:sim5} and Figure~\ref{fg:sim6} depict the worst-case intrusion detectability versus the number of radar receivers for $3$, $5$, and $10$ radar transmitters, respectively, where we also plot the shortcut barrier-based optimal deployment scheme (OPT). The results for RAN are averaged over 100 simulation runs. We observe that OPT performs significantly better than RAN. The reason is that deployment on a barrier is essentially much more efficient than deployment in the full field for worst-case coverage.

\section{Related Work}\label{sc:rlt}

Worst-case coverage has been first studied in \cite{Meguerdichian01INFOCOM} for a traditional sensor network. Polynomial-time algorithms are devised to find maximum breach and maximum support paths between two locations. An efficient algorithm is proposed in
\cite{Li03TC} to solve the best-coverage problem raised in \cite{Meguerdichian01INFOCOM}. In \cite{Meguerdichian01MOBICOM},
efficient algorithms are developed to find the minimum exposure path in sensor networks. Localized algorithms are designed in
\cite{Meguerdichian01MOBIHOC} to solve the minimum exposure path problem. A new coverage measure that captures both the best
and worst-case coverage is studied in \cite{Lee10}. The deployment problem to improve the maximal breach path is considered
by \cite{Gau06,Duttagupta06}.

Barrier coverage is an intimately related problem to worst-case coverage. The concept of weak and strong barrier coverage has been introduced in \cite{Kumar05}, where critical condition of weak barrier coverage is obtained for random deployment. The critical condition of strong barrier coverage is derived in \cite{Liu08} using percolation theory. An effective metric of barrier coverage quality is proposed in \cite{Chen08}. \cite{Saipulla10} studies constructing barrier by sensors with limited mobility after initial deployment. A novel full-view coverage model is proposed in \cite{Wang11} for constructing barrier in camera sensor networks.

\section{Conclusion}\label{sc:ccl}

Radar technology has great potential in many applications, such as border monitoring, security surveillance. In this paper, we studied the worst-case coverage for a bistatic radar network consisting of multiple radar transmitters and radar receivers, where each pair of radar transmitter and
receiver can form a bistatic radar. The problem of optimal radar deployment is highly non-trivial since 1) the detection range of a bistatic radar is characterized by Cassini oval which presents complex geometry; 2) the detection ranges of different bistatic radars are coupled and the network coverage is intimately related to the locations of all radar nodes. We present a general assumption on the field geometry under which it is optimal to deploy radars on a shortest line segment across the field, for maximizing the worst-case intrusion detectability. Further, we characterized the corresponding optimal deployment locations along this shortest line segment. Specifically, the optimal deployment locations exhibits a balanced structure. We also developed a polynomial-time approximation algorithm for characterizing the worst-case intrusion path for any given deployment of radars.

To the best of our knowledge, the optimality of line-based deployment for the worst-case coverage, in particular for bistatic radar networks, has not been studied before this work. Although the detectability model involves some idealized assumptions, we believe that this work will be of value in setting the foundations for networked radar systems. There are still many questions remaining open for the design of networked radar.

\bibliographystyle{IEEEtran}
\bibliography{IEEEabrv,mobihoc_Bib}

\section*{APPENDIX}

\subsection*{A. PROOF OF THEOREM~\ref{thm:mbp}}

\begin{figure}[t]
\vspace{-1cm}
\centering
\includegraphics[width=0.40\textwidth]{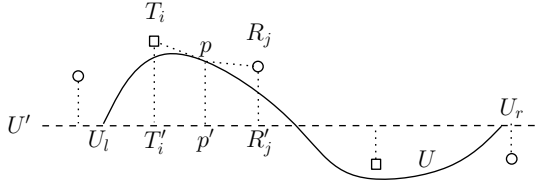}
\caption{An arbitrary radar deployment forms a curve barrier $U$ (solid curve), while
a line-based radar deployment can form a line barrier $\overline{U_lU_r}$ (dashed line) with no greater vulnerability.} \label{fg:proj}
\end{figure}

\textit{Proof:} Consider any given deployment locations $\{\mathcal{T}, \mathcal{R}\}$. It suffices to show that there exist
deployment locations $\{\mathcal{T}', \mathcal{R}'\}$ on $H$ such that $B(P^*) \ge B'({P^*}')$, where $'$ indicates
that the deployment locations are $\{\mathcal{T}', \mathcal{R}'\}$. It follows from Property~1 that there exists a curve barrier
$U$ with $U\subset A(B(P^*))$. Let $U_l$ and $U_r$ be the end points of curve $U$ with $U_l\in F_l$ and $U_r\in F_r$. We will show that the deployment at $\{\mathcal{T}', \mathcal{R}'\}$ forms a line barrier $\overline{U_lU_r}$ with
$Q'(\overline{U_lU_r}) \le Q(U)$ (see Figure~\ref{fg:proj}).

Let $U'$ be the line passing through $U_l$ and $U_r$. Then we can find the projection of any $T_i \in \mathcal{T}$ and $R_j
\in \mathcal{R}$ onto line $U'$, denoted by $T'_i$ and $R'_j$, such that $\overline{T_iT'_i} \perp U'$ and $\overline{R_jR'_j}
\perp U'$. Consider any $p'\in \overline{U_lU_r}$. Since curve $U$ is a path from $U_l$ to $U_r$, we can find a $p\in U$ such
that $p'$ is the projection of $p$ onto line $U'$. Then it follows from the property of projection that $\|T_ip\| \ge
\|T'_ip'\|$ and $\|R_jp\| \ge \|R'_jp'\|$ for all $i$, $j$. Hence, from~\eqref{det} we have $I(p) \le I'(p')$, and it follows that $Q'(\overline{U_lU_r}) \le Q(U)$ due to \eqref{vul}. Since $H$ is the shortcut barrier, we have $h \le \|U_lU_r\|$. Thus, we can translate the line-based deployment $\{\mathcal{T}', \mathcal{R}'\}$ onto the line where $H$ lies such that $Q'(H) \le Q(U)$. Note that $\{\mathcal{T}', \mathcal{R}'\}$ may not be all on $H$. In this case, we can move each $T'_i\notin H$ or $R'_j\notin H$ to its closest end point of $H$ such that $I'(p)$ is not increased for all $p\in H$ and hence $Q'(H) \le Q(U)$ still holds. Therefore, since $H$ is a barrier and $U \subset A(B(P^*))$, we have $B'({P^*}') \le Q'(H) \le Q(U) \le B(P^*)$.
$\hfill\square$

\subsection*{B. PROOF OF THEOREM~\ref{thm:bl}}

\begin{figure}[t]
\centering
\includegraphics[width=0.25\textwidth]{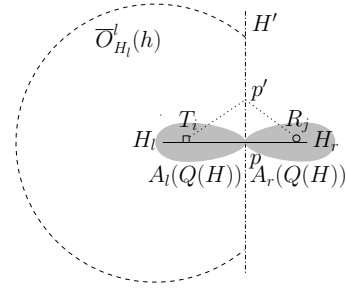}
\caption{Any barrier with possibly lower vulnerability than the shortcut barrier $H$ must lie in the detection range $A(Q(H))$ (grey area)
and pass through a line $H'$ which has a unique intersection $p$ with $A(Q(H))$.} \label{fg:circle}
\end{figure}

\textit{Proof:} Consider any given deployment locations $\{\mathcal{T}, \mathcal{R}\}$ on $H$. From the definition of barrier we have $B(P^*) \le Q(H)$. Suppose $B(P^*) < Q(H)$. The proof is based on contradiction. It follows from Property~1 that there
exists a barrier $U$ with $U\subset A(B(P^*)) \subset A(Q(H))$. We will show that barrier $U$ must intersect with
$H$ (see Figure~\ref{fg:circle}).

\textit{Step 1:} Consider a point $p\in H$ with $I(p) = Q(H)$. Suppose $p\neq H_l$ and $p\neq H_r$. Let $H'$ be the line passing through $p$ such that $H'\perp H$. Consider any $p' \in H'$ with $p' \neq p$. Since $\|T_ip'\| > \|T_ip\|$ and $\|R_jp'\| > \|R_jp\|$ for all $i,j$, we have $I(p') > I(p)$ and hence $p' \notin A(Q(H))$. Thus $H'$ intersects with $A(Q(H))$ at a unique point $p$.

\textit{Step 2:} Let $A_l(Q(H))$ and $A_r(Q(H))$ denote the subsets of $A(Q(H))$ on the left and right side of $H'$,
respectively (see Figure~\ref{fg:circle}). Define $O_c(r)$ as an open disk centered at point $c$ with radius $r$. We will show
that $A_l(Q(H))\subset O_{H_l}(h)$ and $A_r(Q(H))\subset O_{H_r}(h)$.

Let $\overline{O}_{H_l}(h)$ denote the circle centered at $H_l$ with radius $h$. Then let $\overline{O}^l_{H_l}(h)$ denote the subset of
$O_{H_l}(h)$ on the left side of $H'$, which is an arc. It can be verified that $\|T_ia\| > \|T_ip\|$ and $\|R_ja\| >
\|R_jp\|$ for all $i,j$ and for all $a\in \overline{O}^l_{H_l}(h)$. Then since $I(p) = Q(H)$, we have $A(Q(H))\cap \overline{O}^l_{H_l}(h) =
\varnothing$. This implies that $A_l(Q(H))\subset O_{H_l}(h)$. Similarly, we can show $A_r(Q(H))\subset O_{H_r}(h)$.

\textit{Step 3:} Let $U_l$ and $U_r$ be the end points of $U$ with $U_l\in F_l$ and $U_r\in F_r$. Then we have $U_l\notin O_{H_r}(h)$ and $U_r\notin O_{H_l}(h)$, since otherwise $\|U_lH_r\|< h$ or $\|H_lU_r\|< h$, which contradicts that $H$ is the shortcut barrier. Using the result of Step 2, we have $U_l\in A_l(Q(H))$ and $U_r\in A_r(Q(H))$. Thus, curve $U$ must intersect $H'$. From the result of Step 1, $U$ must pass through $p$. This implies that $Q(U) \ge I(p) = Q(H)$, which contradicts that $Q(U) \le B(P^*) < Q(H)$.
$\hfill\square$

\subsection*{C. PROOF OF LEMMA~\ref{lm:local}}


We start with an observations which is used repeatedly in this proof. Clearly, for any $p\in H$, $I(p)$ only depends on the
distances of $p$ to its closest transmitter and receiver, say $T_i$ and $R_j$, respectively. If $p\notin \overline{T_iR_j}$,
$I(p) = \|T_ip\|\|R_jp\|$ decreases as $p$ moves towards $T_i$ and $R_j$; if $p\in \overline{T_iR_j}$, since $\|T_ip\| +
\|pR_j\| = \|T_iR_j\|$ is a constant, it can be easily verified that $I(p) = \|T_ip\|\|pR_j\|$ increases as $p$ moves towards
$Y_{T_iR_j}$, and hence attains maximum when $p = Y_{T_iR_j}$.

The main idea of this proof is to divide the line segment between each pair of neighbor nodes into intervals such that all
points on an interval have the same pair of closest transmitter and receiver, and then we examine the structure of
detectability on each interval using the observation presented above. Due to space limitation, we only prove the case when $S_k$ and
$S_{k+1}$ have different types (transmitter type and receiver type) for some $k$. The same idea is used to prove other cases.

%
%

Suppose $S_k$ and $S_{k+1}$ have different types. Without loss of generality (WLOG), let $S_k = T_i$ and $S_{k+1} = R_j$. 1)
Suppose the closest transmitter and receiver are $T_i$ and $R_j$ for any $p\in \overline{T_iR_j}$. Since $I(p) =
\|T_ip\|\|pR_j\|$ for $p\in \overline{T_iR_j}$, $I(p)$ increases as $p$ moves towards $Y_{T_iR_j}$. 2) Suppose there exists
some $R_m\in \overline{H_lT_i}$ such that $Y_{R_mR_j}\in \overline{T_iY_{T_iR_j}}$. Then $I(p) = \|R_mp\|\|T_ip\|$ for $p\in
\overline{T_iY_{R_mR_j}}$, and $I(p) = \|T_ip\|\|pR_j\|$ for $p\in \overline{Y_{R_mR_j}Y_{T_iR_j}}$. For each case of $p$,
$I(p)$ increases as $p$ moves towards $Y_{T_iR_j}$. Similarly, if there exists some $T_n\in \overline{R_jH_r}$ such that
$Y_{T_iT_n}\in \overline{Y_{T_iR_j}R_j}$, we can also show that $I(p)$ increases as $p$ moves towards $Y_{T_iR_j}$ for $p\in
\overline{Y_{T_iR_j}R_j}$.


\subsection*{D. PROOF OF LEMMA~\ref{lm:bal_loc}}

Due to space limitation, we only prove the case of $\mathbf{S}_i$ with pattern $\mathbf{P}_1$, and the same idea is used to prove other
cases. The main idea is to determine the spacing between two neighbor nodes successively according to the given order of
nodes.

Suppose $\mathbf{S}_i = (T_1, R_1,\cdots, R_k, H_r)$ and $\mathbf{D}_{\mathbf{S}_i}$ is balanced with $Q(\overline{T_1H_r}) =
c$. It follows from $I(Y_{T_1R_1}) = (\|T_1R_1\|/2)^2$ $= c$ that $\|T_1R_1\| = 2\sqrt{c}$. Since $I(Y_{R_1R_2}) =$
$\|T_1Y_{R_1R_2}\|$ $\|Y_{R_1R_2}R_2\| =$ $(\|T_1R_1\| +$ $\|R_1R_2\|/2)$ $\\(\|R_1R_2\|/2) = c$, given $\|T_1R_1\|$
$=2\sqrt{c}$, we obtain a unique value of $\|R_1R_2\|$. Similarly, in a recursive manner, given the values of $\|T_1R_1\|$,
$\cdots$, $\|R_{i-1}R_i\|$, we obtain a unique value of $\|R_iR_{i+1}\|$ such that $I(Y_{R_iR_{i+1}}) = c$ $\ldots$ until we
obtain a unique value of $\|R_kH_r\|$ such that $I(H_r) = c$. Then we can see that $\|T_1R_1\| = e_0(c)$, $\|R_1R_2\| =
e_1(c)$, $\cdots$, $\|R_{k-1}R_k\| = e_{k-1}(c)$, $\|R_kH_r\| = e_k(c)/2$. Similar results can be shown for any $\mathbf{S}_i$
with pattern $\mathbf{P}_1$.


\subsection*{E. PROOF OF LEMMA~\ref{lm:bal_loc_opt}}

Due to space limitation, we only prove the case of $\mathbf{S}_i$ with pattern $\mathbf{P}_1$ and the same idea is used to prove other
cases. Suppose $\mathbf{S}_i = (T_1,R_1,\cdots,R_k,H_r)$ and $\mathbf{S}'_i = (T'_1,R'_1, \cdots,$ $R'_k, H'_r)$. Let $I'(p)$
denote the detectability of $p$ under $\mathbf{S}'_i$. The proof is based on contradiction. Suppose $\|T'_1H'_r\| >
\|T_1H_r\|$. Since $I(Y_{T_1R_1}) = (\|T_1R_1\|/2)^2 = c \ge I'(Y_{T'_1R'_1}) = (\|T'_1R'_1\|/2)^2$, we have $\|T'_1R'_1\| \le
\|T_1R_1\|$. Using this, we can show $\|T'_1R'_2\| \linebreak \le \|T_1R_2\|$. Then following a similar argument, using $\|T'_1R'_2\|
\le \|T_1R_2\|$, we can show $\|T'_1R'_3\| \le \|T_1R_3\|$ $\ldots$ until $\|T'_1R'_k\| \le \|T_1R_k\|$.


Since we have supposed $\|T'_1H'_r\| > \|T_1H_r\|$, we must have $\|R'_kH'_r\| = \|T'_1H'_r\| - \|T'_1R'_k\| > \|T_1H_r\| -
\|T_1R_k\| = \|R_kH_r\|$. It follows that $I'(H'_r) = \|T'_1H'_r\|\|R'_kH'_r\| > \|T_1H_r\|\|R_kH_r\| = c$, which is a
contradiction. Thus we conclude $\|T'_1H'_r\| \le \|T_1H_r\|$. Further, it can be easily verified that if $\|T'_1H'_r\| =
\|T_1H_r\|$, we have $\|T'_1R'_1\| = \|T_1R_1\|$, $\cdots$, $\|R'_kH'_r\| = \|R_kH_r\|$. Similar results can be shown for any
$\mathbf{S}_i$ with pattern $\mathbf{P}_1$.


\subsection*{F. PROOF OF LEMMA~\ref{lm:swap}}

Suppose $\mathbf{S} = (H_l,\cdots,T_1,T_2,R_1,R_2,\cdots,H_r)$ with any spacings. We can construct a new order of locations
$\mathbf{S}' = (H'_l,\cdots$, $T'_1,R'_1,T'_2,R'_2,\cdots,H'_r)$ from $\mathbf{S}$ by swapping the locations of nodes $T_2$ and
$R_1$. Let $I'(p)$ denote the detectability of $p$ after the swapping. We observe that for any local vulnerable point $p\in
\overline{H_lT_1}$, the closest transmitter is not $T_2$, and for any local vulnerable point $p\in \overline{R_2H_r}$,
$\|T_2p\| \ge \|T'_2p\|$. Similar observations can be made for $R_1$ and $R'_1$. Thus we can see that $I'(p)\le I(p)$ for any
local vulnerable point $p\in \overline{H_lT_1}$ or $p\in \overline{R_2H_r}$. We also observe that $I(Y_{T_1T_2}) \ge
(\|T_1T_2\|/2)^2 = (\|T'_1R'_1\|/2)^2 = I'(Y_{T'_1R'_1})$. Similarly, we can obtain $I(Y_{R_1R_2}) \ge I'(Y_{T'_2R'_2})$.
Furthermore, we have $I(Y_{T_2R_1}) = (\|T_2R_1\|/2)^2 =$ $(\|R'_1T'_2\|/2)^2 = I'(Y_{R'_1T'_2})$. Thus we have shown that
$I'(p) \le I(p)$ for any local vulnerable point $p\in \overline{H_lH_r}$, and hence, $\linebreak Q(\overline{H_lH_r})$ is not increased
after the swapping. Similar results can be shown for an order with any pattern given in the claim by a swapping argument.

\subsection*{G. PROOF OF THEOREM~\ref{thm:bal_opt}}

Suppose $\mathbf{D}_{\mathbf{S}}$ is balanced with $Q(\overline{H_lH_r}) = c$. We have shown that $\mathbf{S}$ consists of a
series of local suborders $\mathbf{S}_1,\cdots,\mathbf{S}_m$. Suppose $\mathbf{S}'$ is the same order as $\mathbf{S}$ and with
spacings $\mathbf{D}_{\mathbf{S}'}$ such that $Q'(\overline{H'_lH'_r}) \le c$, where $Q'(\overline{H'_lH'_r})$ denote the
vulnerability of $\overline{H'_lH'_r}$ under $\mathbf{S}'$. Then $\mathbf{S}'$ also consists of a series of local suborders
$\mathbf{S}'_1,\cdots,\mathbf{S}'_m$ such that $\mathbf{S}_j$ and $\mathbf{S}'_j$ are the same local suborders for all
$j=1,\cdots,m$. It is clear that $Q'(Z_{\mathbf{S}'_j}) \le c$, $\forall j$. Then Lemma~\ref{lm:bal_loc_opt} implies that
$L_{\mathbf{S}'_i} \le L_{\mathbf{S}_j}$, $\forall j$. Since $\sum^m_{j=1}L_{\mathbf{S}_j} = h =
\sum^m_{j=1}L_{\mathbf{S}'_j}$, we must have $L_{\mathbf{S}'_j} = L_{\mathbf{S}_j}$, $\forall j$. Hence, by
Lemma~\ref{lm:bal_loc_opt}, we have $\mathbf{D}_{\mathbf{S}'} = \mathbf{D}_{\mathbf{S}}$.

\subsection*{H. PROOF SKETCH OF THEOREM~\ref{thm:ord_opt}}

Consider any given candidate order $\mathbf{S}$. Clearly, $\mathbf{S}$ consists of suborders $\mathbf{S}_1$, $\cdots$,
$\mathbf{S}_{M+1}$ where $\mathbf{S}_1 = (H_l,\cdots,T_1)$, $\cdots$, $\mathbf{S}_i = (T_{i-1},\cdots,T_i)$, $\cdots$,
$\mathbf{S}_{M+1} = (T_M,\cdots,H_r)$. Suppose $\mathbf{D}_{\mathbf{S}}$ is balanced with $\|H_lH_r\| = h$ and
$Q(\overline{H_lH_r})$ $= c$.

The proof has two steps. In step 1, we suppose the conditions given in the claim do not hold, and then we can find a new order
$\mathbf{S}'$ from $\mathbf{S}$ with some spacings $\mathbf{D}_{\mathbf{S}'}$ such that $Q'(\overline{H'_lH'_r}) \le c$ and
$\|H'_lH'_r\| \ge \|H_lH_r\|$, where $Q'(\overline{H'_lH'_r})$ denote the vulnerability of $\overline{H'_lH'_r}$ under
$\mathbf{S}'$. This implies that $\mathbf{S}'$ is no worse than $\mathbf{S}$. Repeating this argument, we can eventually
find a new order $\mathbf{S}'$ satisfying the conditions in the claim and is no worse than $\mathbf{S}$. In step 2, we
show that all candidate orders satisfying the conditions in the claim are equivalently good. This implies that they are all
optimal orders. Due to space limitation, we only prove step 1 for the case $n_{T_{i-1}T_i} \ge n_{T_{j-1}T_j} + 2$ and
$n_{T_{j-1}T_j} \ge 2$. The same idea is used to prove other cases in step 1 although more intricate arguments are needed for
some cases.

Suppose $n_{T_{i-1}T_i} \ge n_{T_{j-1}T_j} + 2$ and $n_{T_{j-1}T_j} \ge 2$. Clearly, $\mathbf{S}_i$ and $\mathbf{S}_j$ must be
local suborders. We can construct a new order $\mathbf{S}'$ from $\mathbf{S}$ by moving a receiver from between $T_{i-1}$ and
$T_i$ in $\mathbf{S}_i$ to between $T_{j-1}$ and $T_j$ in $\mathbf{S}_j$. Then we set the spacings $\mathbf{D}_{\mathbf{S}'_i}$
and $\mathbf{D}_{\mathbf{S}'_j}$ balanced and $Q(Z_{\mathbf{S}'_i}) = Q(Z_{\mathbf{S}'_j}) = c$, and we keep the spacings of
any other suborder unchanged, i.e., $\mathbf{D}_{\mathbf{S}_m} = \mathbf{D}_{\mathbf{S}'_m}$ for all $m\neq i,j$. Therefore, we
must have $Q'(\overline{H'_lH'_r}) = c$.

We observe from Lemma~\ref{lm:bal_loc} that we have $L_{\mathbf{S}'_j} - L_{\mathbf{S}_j} = e_{(n_{T_{j-1}T_j}+1)/2}(c)$ if
$n_{T_{j-1}T_j}$ is odd or $e_{n_{T_{j-1}T_j}/2}(c)$ if $n_{T_{j-1}T_j}$ is even. Also, we have $L_{\mathbf{S}_i} -
L_{\mathbf{S}'_i} = e_{(n_{T_{i-1}T_i}-1)/2}(c)$ if $n_{T_{i-1}T_i}$ is odd or $e_{n_{T_{i-1}T_i}/2}(c)$ if $n_{T_{i-1}T_i}$ is
even. Since $e_0(c) > e_1(c) > e_2(c) > \cdots$ for any $c$, in any case, we have $\|H'_lH'_r\| - \|H_lH_r\| =
(L_{\mathbf{S}'_j} - L_{\mathbf{S}_j}) - (L_{\mathbf{S}_i} - L_{\mathbf{S}'_i}) > 0$. Recall that we have shown
$Q'(\overline{H'_lH'_r}) = c$. Then the desired result follows.

\end{document}